\documentclass{iau} 
\usepackage{graphicx}
\usepackage{natbib}
\usepackage{url}

\newcommand*\aap{A\&A}

\newcommand*\apj{ApJ}

\newcommand*\mnras{MNRAS}

\title[Synthetic clusters of massive stars] %% give here short title %%
{Synthetic clusters of massive stars to test stellar evolution models}

\author[Cyril Georgy \& Sylvia Ekstr\"om]   %% give here short author list %%
{Cyril Georgy$^{1, 2}$
 \and Sylvia Ekstr\"om$^1$}

\affiliation{$^1$Geneva Observatory, Geneva University\\Chemin des Maillettes 51, 1290 Versoix, Switzerland\\ email: {\tt cyril.georgy@unige.ch} \\ email: {\tt sylvia.ekstrom@unige.ch}\\[\affilskip]
$^2$Astrophysics Group, Keele University \\ Keele, ST5 5BG, United Kingdom}

\pubyear{2015}
\volume{xxx}  %% insert here IAU Symposium No.
\jname{Title of your IAU Symposium}
\editors{A.C. Editor, B.D. Editor \& C.E. Editor, eds.}
\begin{document}

\maketitle

\begin{abstract}
During the last few years, the Geneva stellar evolution group has released new grids of stellar models, including the effect of rotation and with updated physical inputs \citep{Ekstrom2012a,Georgy2013b,Georgy2013a}. To ease the comparison between the outputs of the stellar evolution computations and the observations, a dedicated tool was developed: the \textsc{Syclist} toolbox \citep{Georgy2014b}. It allows to compute interpolated stellar models, isochrones, synthetic clusters, and to simulate the time-evolution of stellar populations.
%% add here a maximum of 10 keywords, to be taken form the file <Keywords.txt>
\end{abstract}

\firstsection % if your document starts with a section,
              % remove some space above using this command.
\section{The \textbf{{\sc{Syclist}}} toolbox}

In cluster mode, \textsc{Syclist}\footnote{\url{http://obswww.unige.ch/Recherche/evoldb/index/}} has (among others) the following capabilities: it accounts for a Salpeter IMF, includes several initial distributions for the angular velocity of the stars, as well as the effect of a random distribution of the angle of view on the gravity darkening. It can mimic in a simplified way the presence of a fraction of binary stars in the population.

The effect of fast rotation on the shape of the star is included in the Roche model approximation. The variation of the effective temperature and of the luminous flux over the surface (the so-called gravity darkening) is implemented using either the \citet{vonZeipel1924a} or \citet{EspinosaLara2011a} relations. In this framework, the equator of a fast rotating star is cooler and dimmer than the average temperature and luminosity, and the poles are hotter and brighter.

As a consequence, an observer ignoring the orientation of the axis of rotation of a star will deduce an erroneous temperature and luminosity depending on its angle of view. For example, an observer looking at a fast rotating star pole-on will see principally the hot and bright polar regions and thus deduce that the star is bluer and more luminous than the average surface values.

\begin{figure}
% \vspace*{-2.0 cm}
\begin{center}
 \includegraphics[width=.85\textwidth]{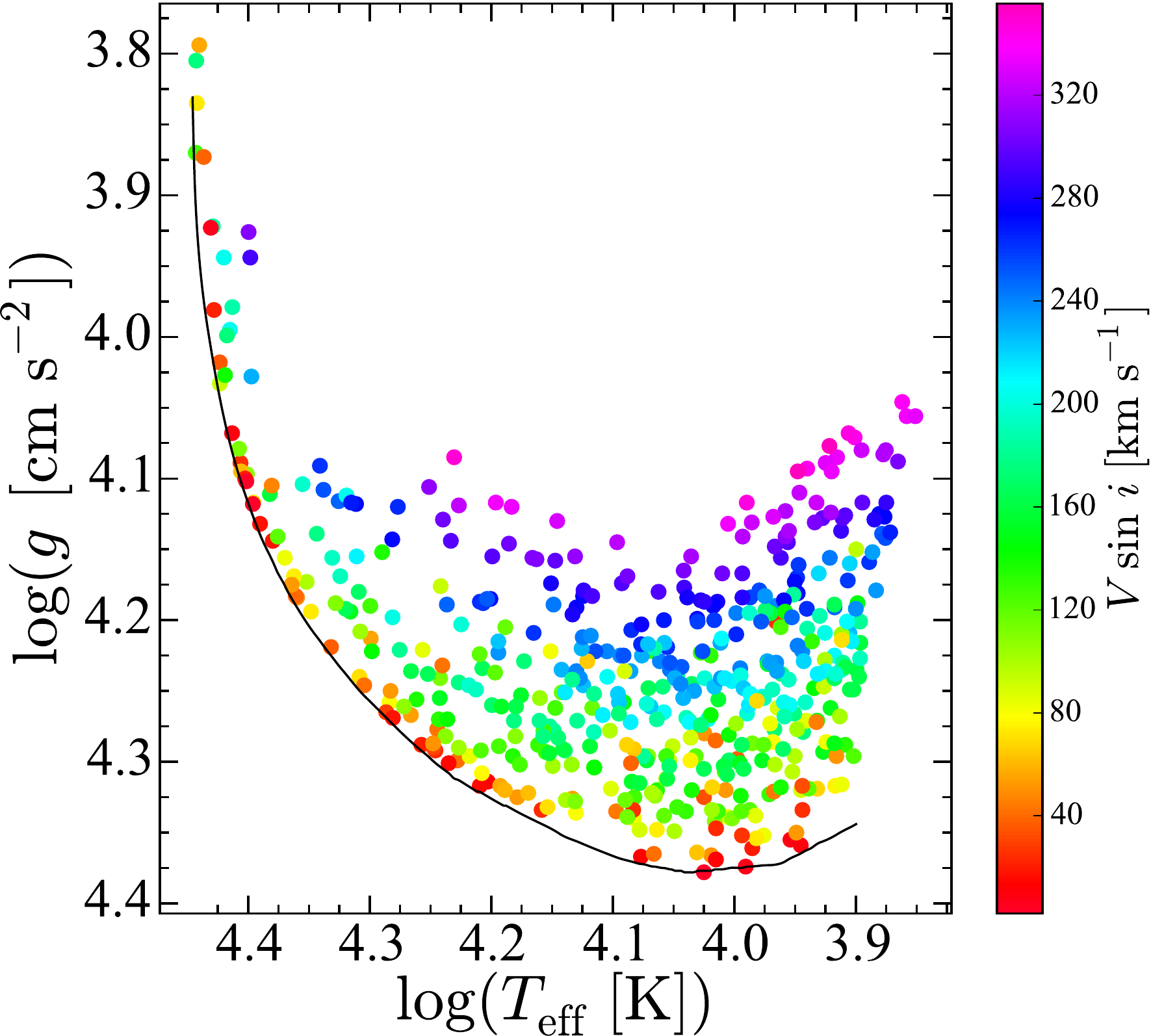} 
% \vspace*{-1.0 cm}
 \caption{`Observed' $g_\mathrm{eff}$ as a function of `observed' $T_\mathrm{eff}$ for a cluster at 25 Myr at $Z_\odot$. The colour code shows the $V\sin(i)$ value. The black line is the corresponding isochrone for a non rotating population. No instrumental noise has been considered here.}
   \label{Fig1}
\end{center}
\end{figure}

The latitudinal variation of the gravity has also observational consequences. Assuming that the gravity deduced from a spectral fitting corresponds to the flux-averaged gravity of the visible hemisphere, a random distribution of the angle of view will produce a scatter in the $g_\mathrm{eff} -T_\mathrm{eff}$ plane (Fig.~\ref{Fig1}).

Extended main-sequence turn-off in the colour-magnitude diagram is a common feature of star clusters. One explanation is that these clusters, instead of having an instantaneous burst of star formation, have a spread \citep[e.g.][]{Goudfrooij2014a,Correnti2014a}. Recently, we have shown that rotation could be a plausible alternative explanation, providing a natural framework to the observed relation between the duration of the spread in the star formation rate and the age of the cluster \citep{Niederhofer2015a}. In this framework, the extension of the turn-off is no more produced by a spread, but by the distribution of initial rotation rates of the stars.

\acknowledgments{CG acknowledges support from the European Research Council under the European Union's Seventh Framework Programme (FP/2007-2013) / ERC Grant Agreement n. 306901.}

\end{document}